\documentstyle[psfig,emulateapj]{article}

\begin{document}

\title{Resolving the submillimeter background: the 850 micron galaxy counts}
\author{A.~J.~Barger,\altaffilmark{1}
L.~L.~Cowie,\altaffilmark{1}
D.~B.~Sanders\altaffilmark{1}
}

\altaffiltext{1}{Institute for Astronomy, University of Hawaii, 
2680 Woodlawn Drive, Honolulu, Hawaii 96822, USA}

\slugcomment{Accepted by The Astrophysical Journal Letters}

\begin{abstract}
Recent deep blank field submillimeter (submm) surveys have revealed
a population of luminous high redshift galaxies that emit most of their 
energy in the submm. The results suggest that much of the
star formation at high redshift may be hidden to optical observations.
In this paper we present wide-area 850-$\mu$m SCUBA data on 
the Hawaii Survey Fields SSA13, SSA17, and SSA22. Combining these new data 
with our previous deep field data, we establish the 
850-$\mu$m galaxy counts from 2\ mJy
to 10\ mJy with a $>3\sigma$ detection limit. 
The area coverage is 104\ arcmin$^2$ to 8\ mJy
and 7.7\ arcmin$^2$ to 2.3\ mJy.
The differential 850-$\mu$m counts are well described by the function
$n(S)=N_0/(a+S^{3.2})$, where $S$ is the flux in mJy, 
$N_0=3.0\times 10^4$ per square degree per mJy, and $a=0.4-1.0$ is chosen
to match the 850-$\mu$m extragalactic background light.
Between 20 to 30\ per cent of the 850-$\mu$m background 
resides in sources brighter than 2\ mJy. Using an empirical fit to 
our $>2$\ mJy data constrained by the EBL at lower fluxes, 
we argue that the bulk of
the 850-$\mu$m extragalactic background light resides in sources with
fluxes near 1\ mJy. The submm sources are plausible progenitors of the 
present-day spheroidal population.

\end{abstract}


\section{Introduction}

The past few years have seen a dramatic evolution in our knowledge of the
history of star formation in optically-selected galaxies covering
the redshift range from $z=0$ to $z\sim 6$ (see, e.g., 
\markcite{madau98}Madau, Pozzetti, \& Dickinson 1998 and references therein;
\markcite{dey98}Dey et al.\ 1998; 
\markcite{hu98}Hu, Cowie, \& McMahon 1998; 
\markcite{weymann98}Weymann et al.\ 1998;
\markcite{cowie99}Cowie, Songaila, \& Barger 1999).
However, it has long been suspected that much of the
early star formation in galaxies takes place inside shrouds of dust
(e.g.\ \markcite{pp67}Partridge \& Peebles 1967). 
While it is possible to attempt to correct the optical star formation 
history for the effects of dust (\markcite{heckman98}Heckman et al.\ 1998; 
\markcite{pettini98}Pettini et al.\ 1998), 
the corrections are far from certain.
However, the importance of absorption and reradiation of light by dust 
in the history of galaxy formation and evolution is evident: the
submm extragalactic background light (EBL)
(\markcite{puget96}Puget et al.\ 1996; 
\markcite{guider97}Guiderdoni et al.\ 1997;
\markcite{schlegel98}Schlegel et al.\ 1998;
\markcite{fixsen98}Fixsen et al.\ 1998; 
\markcite{hauser98}Hauser et al.\ 1998)
has approximately the same integrated energy density as the optical EBL.
Thus, an accurate reconstruction of the global star formation history 
versus redshift requires an understanding of both the 
rest-frame ultraviolet and the rest-frame far-infrared properties of the 
galaxy populations.

Reradiation of stellar light by dust into the
far-infrared produces a thermal emission peak at $\lambda\sim 60-100$-$\mu$m
that is redshifted into the submm for galaxies at $z>1$.
Submm observations are unique in that the strong negative
K-corrections at 850-$\mu$m nearly compensate for
cosmological dimming beyond $z\simeq 1$, thereby making dusty galaxies
almost as easy to detect at $z\simeq 10$ as at $z\simeq 1$
(\markcite{bl93}Blain \& Longair 1993; 
\markcite{mcmahon94}McMahon et al.\ 1994).
The revolutionary new camera, SCUBA (Submillimeter Common User 
Bolometer Array; \markcite{holland98}Holland et al.\ 1998), 
on the 15-m James Clerk Maxwell Telescope 
(JCMT)\altaffilmark{2}\altaffiltext{2}{The JCMT is operated by 
the Joint Astronomy Centre on behalf of the parent organizations
the Particle Physics and Astronomy Research Council in the United
Kingdom, the National Research Council of Canada, and The Netherlands
Organization for Scientific Research.} 
on Mauna Kea resolves the submm EBL into its individual components
through a combination of unparalleled sensitivity and large field-of-view.
Recent SCUBA surveys
(\markcite{smail97}Smail, Ivison, \& Blain 1997;
\markcite{smail98}Smail et al.\ 1998;
\markcite{barger98}Barger et al.\ 1998; 
\markcite{hughes98}Hughes et al.\ 1998; 
\markcite{eales99}Eales et al.\ 1999;
\markcite{lilly99}Lilly et al.\ 1999;
\markcite{blain99}Blain et al.\ 1999)
have uncovered a substantial population of dusty galaxies
with properties similar to those expected for the distant counterparts to
the most luminous, merging systems observed locally, the ultraluminous
infrared galaxies (ULIGs) (\markcite{sanders96}Sanders \& Mirabel 1996). 
In this paper we present our extensive new wide-area SCUBA surveys
of blank regions of sky. We analyze our source counts both empirically
and in terms of semi-analytic models.

\begin{deluxetable}{lcrrccrccc}
\tablewidth{450pt}
\small
\tablenum{1}
\tablecaption{Source Catalog\label{tab1}}
\tablehead{
\colhead{Name} & \multicolumn{3}{c}{RA(2000)} &
\multicolumn{3}{c}{Dec(2000)} & \colhead{$S_{850\mu{\rm m}}$\ (mJy)} &
\colhead{$N_{850\mu{\rm m}}$\ (mJy)} & \colhead{S/N}
}
\startdata

Lockman Hole & 10 & 34 & 2.1 & 57 & 46 & 25 & 5.1 & 0.69 & 7.4 \\
& 10 & 33 & 56.5 & 57 & 47 & 32 & 2.7 & 0.80 & 3.3 \\
SSA13-deep & 13 & 12 & 32.1 & 42 & 44 & 30 & 3.8 & 0.80 & 4.7 \\
& 13 & 12 & 28.0 & 42 & 44 & 58 & 2.3 & 0.61 & 3.8 \\
& 13 & 12 & 25.7 & 42 & 43 & 50 & 2.4 & 0.74 & 3.2 \\
& 13 & 12 & 28.9 & 42 & 45 & 59 & 3.5 & 1.06 & 3.3 \\
SSA13-wide & 13 & 12 & 19.8 & 42 & 38 & 28 & 5.0 & 1.51 & 3.3 \\
& 13 & 12 & 24.9 & 42 & 39 & 56 & 6.1 & 1.83 & 3.4 \\
& 13 & 12 & 26.3 & 42 & 41 & 46 & 6.3 & 1.91 & 3.3 \\
& 13 & 11 & 59.9 & 42 & 40 & 30 & 10.3 & 2.97 & 3.5 \\
& 13 & 12 & 5.1 & 42 & 44 & 34 & 7.8 & 2.27 & 3.4 \\
SSA17 & 17 & 6 & 25.0 & 43 & 57 & 39 & 8.7 & 2.10 & 4.2 \\
& 17 & 6 & 26.0 & 43 & 54 & 40 & 5.2 & 1.32 & 3.9 \\
& 17 & 6 & 37.2 & 43 & 55 & 31 & 4.8 & 1.29 & 3.7 \\
& 17 & 6 & 33.1 & 43 & 54 & 8 & 5.3 & 1.47 & 3.6 \\
& 17 & 6 & 20.3 & 43 & 54 & 4 & 9.6 & 3.09 & 3.1 \\
SSA22 & 22 & 17 & 34.1 & 00 & 13 & 54 & 8.2 & 1.19 & 6.9 \\
& 22 & 17 & 35.2 & 00 & 15 & 32 & 4.9 & 0.93 & 5.3 \\
& 22 & 17 & 36.1 & 00 & 15 & 54 & 4.0 & 1.02 & 4.0 \\
& 22 & 17 & 34.0 & 00 & 15 & 42 & 3.3 & 0.92 & 3.6 \\
& 22 & 17 & 42.0 & 00 & 16 & 2 & 5.0 & 1.62 & 3.1 \\
\enddata
\end{deluxetable}

\section{Observations}
\label{sec:obs}

In \markcite{barger98}Barger et al.\ (1998) we presented our deepest two 
SCUBA maps of the Lockman Hole and the Hawaii Deep Survey field SSA13,
the observational details of which can be found in that paper. We
note, however, that due to a bug in the SCUBA observing software
during the time these data were taken (discovered and fixed
27 July 1998), the chopper was started
at the updated (Az,El) components corresponding to the requested (RA,Dec) 
chop throw at the beginning of each new 30\ min integration but then 
did not update for the remainder of the integration. Consequently, the 
off-beams appear smeared in arclets from the initial positions. 

Our wide-area SCUBA maps of the Hawaii Deep Survey fields
SSA17 and SSA22 were taken as mosaics during two runs in August 1998
and September 1998 and also during an earlier observing shift in 
October 1997 (totaling 16 observing shifts).
Our wide-area map of SSA13 was taken during a run in February 1999
(7 observing shifts).
The maps were dithered to prevent any regions of the sky from repeatedly 
falling on bad bolometers. The chop throw during the October 1997 run 
was azimuthal. During the runs in 1998 and 1999 the chop throw was fixed at a 
position angle of 90\ deg so that the negative beams would appear 
45\ arcsec on either side East-West of the positive beam.
The negative images can be restored, thereby increasing the effective
exposure times on most of the field.  Every few hours we did a `skydip' 
(\markcite{manual}Lightfoot et al.\ 1998) to measure the zenith
atmospheric opacities at 450 and 850-$\mu$m, and we monitored the
225\ GHz sky opacity at all times to check for sky stability.
Fully sampled wavelength maps at 850-$\mu$m were completed
at median optical depths of 0.38 during 
the SSA17 and SSA22 runs and 0.15 during the SSA13 run.
We performed pointing checks every hour during the observations 
on the blazars 1633+382, 2223-052, or 1308+326. 
We calibrated our data maps using 
30\ arcsec diameter aperture measurements of the positive beam in
once or twice-nightly beam maps of the primary calibration sources Uranus
and Mars.

We reduced all our data in a standard
and consistent way using the dedicated SCUBA User Reduction Facility
(SURF; \markcite{surf}Jenness \& Lightfoot 1998). 
Due to the variation in the density of bolometer samples across the maps,
there is a rapid increase in the noise levels at the very edges;
thus, we clipped the low exposure edges from our images. Our wide-area SCUBA 
maps and multiwavelength follow-up data will be presented elsewhere 
(Barger \& Cowie, in preparation).

%
%

\begin{figure*}[tbh]
\centerline{\psfig{figure=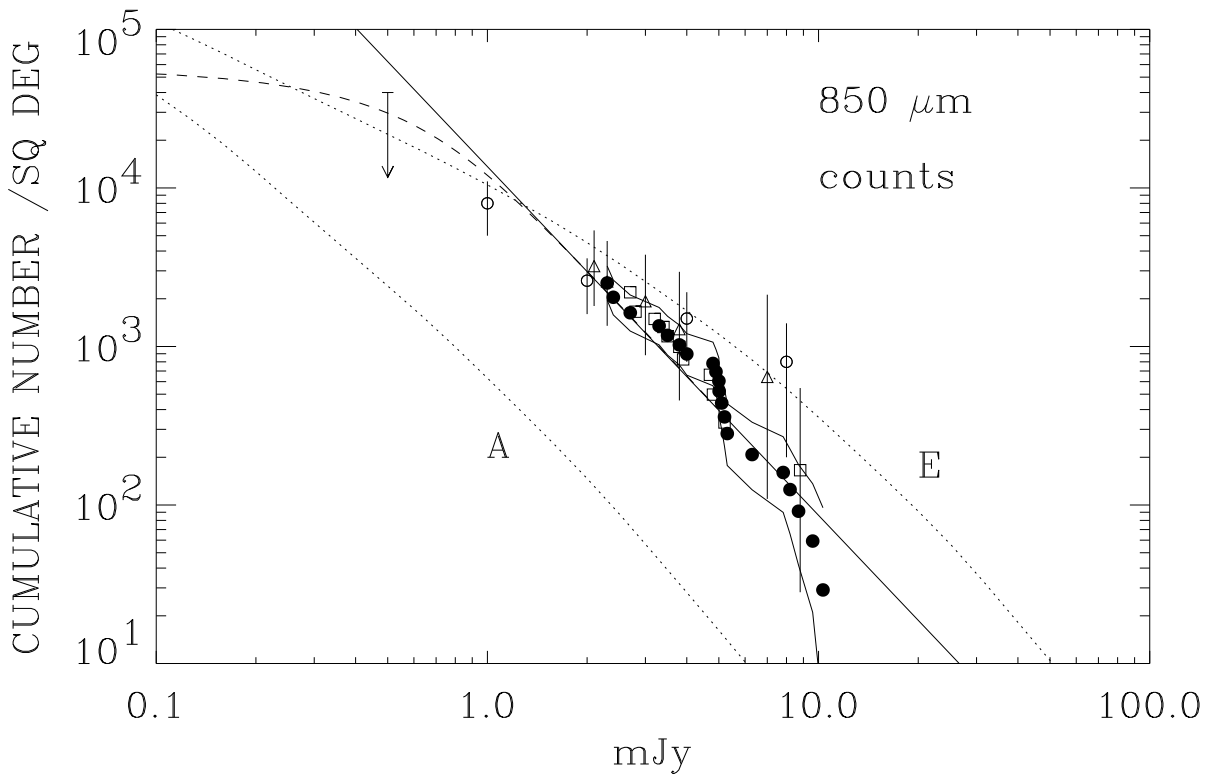,angle=90,width=6.5in}}
\figurenum{1}
\figcaption[barger.fig1.ps]{
Our 850-$\mu$m source counts (solid squares) with $1\sigma$ error limits
(jagged solid lines) are well described by the power-law
parameterization in Eq.~1 with $a=0$, $\alpha=3.2$, and
$N_0=3.0\times 10^4$\ deg$^{-2}$\ mJy$^{-1}$ (solid line).
The dashed curve shows a smooth extrapolation of our fit to EBL measurements
using the value $a=0.5$. Counts from Blain et al.\ (1999)
(open circles), Hughes et al.\ (1998) (open triangles), and
Eales et al.\ (1999) (open squares) are in good agreement with our data
and the empirical fit. Semi-analytic models A and E from Guiderdoni et al.\
(1998) are shown by the dotted curves.
\label{fig1}}
\end{figure*}

\section{Source Extraction and Completeness}
\label{extract}

The SURF reduction routines arbitrarily normalize all the data maps in 
a reduction sequence to the central pixel of the first map; thus, the
noise levels in a combined image are determined relative to the 
quality of the central pixel in the first map. To find the
absolute noise levels, we first eliminated the $\ge 3\sigma$
real sources in each field through subtraction of an appropriately 
normalized version of the beam profile. We then iteratively adjusted
the noise normalization
until the dispersion of the signal to noise ratio
measured at random positions was approximately one.  
This procedure should provide a conservative noise estimate since 
it includes both fainter sources and correlated noise.

We scanned the maps at an array of positions separated by
6$''$ to locate all $\ge 2.8\sigma$ (to overselect) sources. 
We note that the position determinations in the deep maps are not
affected by off-beam smearing because it is a symmetrical effect.
Using the peaked-up positions, we extracted only the $\ge 3\sigma$ sources.
We subtracted the brighter sources from the maps before extracting the 
fainter sources. We calibrated the extracted fluxes, including both positive 
and negative beams, to the 30\ arcsec diameter aperture fluxes of the 
brightest sources; the measured fluxes are therefore unaffected by the 
off-beam smearing.

Our source catalog is given in Table~\ref{tab1}. The noise levels for the 
two deep maps are slightly lower than our initial analysis 
(\markcite{barger98}Barger et al.\ 1998) because of improvements in 
both the SURF reduction routines and our own extraction routines.
These improvements result in an increase in the number of measured 
sources for these fields over our previously published results.

To determine the completeness of our source recovery, we assumed the shape 
of the \markcite{guider98}Guiderdoni et al.\ (1998) `Scenario E' model 
(see \S\ref{disc}) for the counts and simulated the expected sources
by adding appropriately renormalized calibration sources at random
positions to the final maps with the real sources removed.
We then reran our extraction procedure on these images.
Any sources detected at the $\ge 3\sigma$ level were considered 
to be recovered. 
Using this criterion, the output counts matched the
input counts; thus, no incompleteness correction is required.
We also investigated the completeness using the functional form in Eq.~1
(below) that flattens just below the 2\ mJy region and obtained the
same result.
At the $3\sigma$ level, noise may migrate sources above and below
the threshold, but empirically we have found that the net result,
given the shape of the counts versus flux, 
is to leave the overall count determinations unchanged.

Because of non-uniformity across the maps, the flux sensitivity levels and 
area coverages are related. For example, at $3\sigma$ limiting flux levels 
of 20, 10, 8, 5, 4, 3, and 2.3\ mJy, our area coverages, after combining all 
fields, are 141, 122, 104, 43, 31, 18, and 7.7\ arcmin$^2$, respectively. For 
comparison, at a 2.1\ mJy $4\sigma$ flux limit, Hughes et al.\ (1998) have 
an area coverage of 5.6\ arcmin$^2$.
The combined data of Eales et al.\ (1999) at $3\sigma$ flux limits of 
2.8, 2.3, and 1.7\ mJy have areas 21.7, 6.7, and 3.4\ arcmin$^2$,
respectively (private communication from S.~Eales and S.~Lilly); the actual
$1\sigma$ measurement uncertainties given for the 10 and 14-hour sources
in Table~1 of Eales et al.\ (1999) are mostly in the range 0.9--1.1\ mJy.

We present our cumulative 850-$\mu$m source counts per square deg (solid
squares) with the $1\sigma$ errors (jagged solid lines) in Fig.~1.
The cumulative counts are the sum of the inverse areas of all the sources
brighter than flux $S$. We can model the differential counts
as a function of flux in a phenomenological way by fitting
\begin{equation}
n(S)=N_0/(a+S^{\alpha})
\end{equation}
to the data, where $S$ is measured in\ mJy.
Because the data are restricted to the
$>2$\ mJy range, the $a$ parameter makes essentially no difference 
to a fit of observed number versus flux, and we take $a=0$ here.
A chi-square fit gives an optimal index $\alpha=3.2$, with
a 95\ per cent confidence range from 2.6 to 3.9, and a normalization
$N_0=3.0\times 10^4\ {\rm deg}^{-2}\ {\rm mJy}^{-1}$. 
This fit, illustrated in Fig.~1 with a straight solid line, is 
consistent with the absence of $\gg 10$\ mJy sources in our sample.

Over the measured $2-10$\ mJy range, the contribution to
the EBL is $9.3\times 10^3$\ mJy\ {\rm deg}$^{-2}$, which corresponds to
20 to 30\ per cent of the total EBL, depending on whether we adopt the
850-$\mu$m EBL measurement of $4.4\times 10^4$\ mJy\ {\rm deg}$^{-2}$ from
\markcite{fixsen98}Fixsen et al.\ (1998) or
$3.1\times 10^4$\ mJy\ {\rm deg}$^{-2}$
from \markcite{puget96}Puget et al.\ (1996).
Assuming only that the above parameterization provides an appropriately 
smooth continuation to fluxes below 2\ mJy,
we can now fit the EBL with a suitable choice of $a$;
for $a$ in the range $0.4-1.0$ the integral of the
differential counts over all fluxes matches the Fixsen et al.\ and 
Puget et al.\ EBL measurements.

The shape for $a=0.5$ is shown by the dashed line in Fig.~1. 
The open circles with error bars show the cumulative counts
per square degree from the gravitationally-lensed sample of
\markcite{blain99}Blain et al.\ (1999). The Blain et al.\
sample has only one source below 2\ mJy (see last two columns of
their Table~1); at 0.5\ mJy we plot their 2$\sigma$ upper limit.
The open triangles with error bars show the blank field data from 
\markcite{hughes98}Hughes et al.\ (1998).
The open squares show the blank field results from
\markcite{eales99}Eales et al.\ (1999); the statistical 
uncertainty is only indicated on their brightest flux point since the
remainder of their data have similar uncertainties to those on our data.
The agreement between all the data sets is excellent and follows the
empirical fit obtained from our data and the EBL constraint.

\section{Discussion and Conclusions}\label{disc}

The IRAS satellite made the first observations of dust emission from
external galaxies when it measured the sky in four bands
between 12 and 100-$\mu$m. A wide variety of infrared luminosities were
observed, from normal spirals to ULIGs. The spectral energy
distributions (SEDs) of ULIGs are found to provide reasonable fits to the
SEDs of submm sources (\markcite{ivison98}Ivison et al.\ 1998).
Later in this discussion we will assume that the SED of the `prototypical' 
ULIG Arp~220, the spectral shape of which can be represented by a 
modified blackbody with emissivity $\lambda^{-1}$ and a dust temperature 
of $T=47$\ K, is appropriate for the submm sources.

It has long been debated whether the dust-enshrouded local ULIGs are
powered by massive bursts of star formation induced by violent
galaxy-galaxy collisions or by AGN activity. A recent mid-infrared 
spectroscopic survey of 15 ULIGs by
\markcite{genzel98}Genzel et al.\ (1998) found that $70-80$\ per cent 
of the sample are dominantly powered by star formation and 
$20-30$\ per cent by a central AGN.
Similarly, spectroscopic follow-up studies of a gravitationally 
lensed submm sample (\markcite{barger99}Barger et al.\ 1999; 
\markcite{ivison98}Ivison et al.\ 1998) indicate that at least 
20\ per cent of the sample show some AGN activity. We assume in the 
following discussion that a substantial fraction of the submm light 
arises from star formation.

A variety of models have been proposed to predict the submm galaxy
population (e.g.\ \markcite{guider98}Guiderdoni et al.\ 1998; 
\markcite{blain98}Blain et al.\ 1999; 
\markcite{eales99}Eales et al.\ 1999; 
\markcite{trentham99}Trentham, Blain, \& Goldader 1999).
\markcite{guider98}Guiderdoni et al.\ (1998) designed a family of 
evolutionary scenarios that ranged from submm source populations due 
solely to the presence of dust in optical galaxies (Scenario A) to
those due a rapidly increasing fraction of ULIGs with redshift (Scenario E).
In Fig.~1 we compare our blank field 850-$\mu$m cumulative
source counts with these Guiderdoni et al.\ models (dotted lines). 
Scenario A
substantially underpredicts the observed 850-$\mu$m source counts, whereas 
Scenario E overpredicts the counts by a smaller factor. 
Because submm fluxes are approximately independent of redshift, the 
predicted number distributions versus flux are primarily dependent on the 
input luminosity functions. A downward renormalization of Scenario E to 
describe our data would necessitate a steepening of the luminosity 
function at the faint end in order to match the submm EBL. 

Because of the unique aspects of the submm counts, it
is also possible to interpret the data in an empirical way. 
With the empirical $a=0.5$ parameterization of
\S~\ref{extract}, we find that 60\ per cent of the 850-$\mu$m background
light arises from sources lying between just 0.5 and 2.0\ mJy.
The mean source flux averaged over the whole population is 0.7\ mJy.
Because of the steep divergence at low 
flux levels of the empirical fit to the number counts, the turnover 
of the $N(>S)$ distribution will 
occur in essentially the same place independent
of the regularization of the divergence by specific $a$-parameter choices;
hence, our conclusion that the dominant counts occur in the 1\ mJy region
is not sensitive to the relative calibration of the EBL and 
SCUBA observations. We estimate the
cumulative source density in the $1-10$\ mJy range to be
$10^4\ {\rm deg}^{-2}$, which is consistent with the measurement
($0.79\pm 0.3)\times 10^4\ {\rm deg}^{-2}$ by Blain et al.\ (1999);
our predicted density for the $0.5-2.0$\ mJy range is
$2.6\times 10^4\ {\rm deg}^{-2}$.

Limited and still rather uncertain spectroscopic data suggest that most 
of the submm sources lie in the $z=1-3$ 
redshift range (\markcite{barger99}Barger et al.\ 1999;
\markcite{lilly99}Lilly et al.\ 1999).
At these redshifts the far-infrared (FIR) luminosity
is approximately independent of the redshift. 
Thus, if we assume an Arp~220-like spectrum with 
$T=47$\ K (e.g.\ \markcite{barger98}Barger et al.\ 1998),
the FIR luminosity of a characteristic $\sim 1$\ mJy source is in the range 
$4-5\times 10^{11}\ {\rm h_{65}^{-2}\ L_\odot}$
for a $q_0=0.5$ cosmology ($7-15\times 10^{11}$ for $q_0=0.02$). 
The FIR luminosity provides a measure of the current star formation rate 
(SFR) of massive stars (\markcite{sy83}Scoville \& Young 1983;
\markcite{tt86}Thronson \& Telesco 1986),
${\rm SFR}\sim 1.5\ \times 10^{-10}\ (L_{FIR}/{\rm L_\odot})\ 
{\rm M_\odot\ yr^{-1}}$; a 1\ mJy source would 
therefore have a star formation rate 
of $\sim 70\ {\rm h_{65}^{-2}\ M_\odot\ yr^{-1}}$\ for $q_0=0.5$, placing
the `typical' submm source at the high end of SFRs in
optically-selected galaxies
(\markcite{pettini98}Pettini et al.\ 1998). If we were to allow the dust 
temperature to go as low as 30\ K, $L_{FIR}$ and the corresponding SFR would
be $\sim 4$ times smaller.

For a cumulative source density of $4.0\times 10^{4}\ {\rm deg}^{-2}$ 
required to reproduce the EBL with 1\ mJy sources 
($\langle N\rangle={\rm EBL}/\langle S\rangle$ with 
$\langle S\rangle\sim 1$\ mJy)
and redshifts in the $1-3$\ range, the average space density is
$5\times 10^{-3}\ {\rm h_{65}^{3}\ Mpc^{-3}}$
for a $q_0=0.5$ cosmology ($10^{-3}$ for $q_0=0.02$).
This space density is rather insensitive to the upper cut-off on the
redshift distribution, dropping by only a factor of $\sim 2$ or 3 
if we extend the volume calculation to $z=5$.
For comparison, the space density of present-day ellipticals
is about $10^{-3}\ {\rm h_{65}^{3}\ Mpc^{-3}}$
(\markcite{marzke94}Marzke et al.\ 1994; 
\markcite{yoshii88}Yoshii \& Takahara 1988).
Within the still substantial uncertainty posed by the dust temperatures, 
the estimated star formation rate from submm sources is
$\sim0.3\ {\rm h_{65}\ M_\odot\ yr^{-1}\ Mpc^{-3}}$
for $q_0=0.5$, which is nearly an order 
of magnitude higher than that observed in the optical, 
$\sim0.04\ {\rm h_{65}\ M_\odot\ yr^{-1}\ Mpc^{-3}}$
(e.g.\ \markcite{steidel}Steidel et al.\ 1999).

Several groups (\markcite{smail98}Smail et al.\ 1998;
\markcite{eales99}Eales et al.\ 1999;
\markcite{lilly99}Lilly et al.\ 1999;
\markcite{trentham99}Trentham, Blain, \& Goldader 1999)
have suggested that the submm sources are associated with
major merger events giving rise to the formation of spheroidal galaxies.
The approximate equality of the optical and submm backgrounds supports this
hypothesis; present-day spheroidal and disk populations have roughly 
comparable amounts of metal density, and thus their formation is expected 
to produce comparable amounts of light (\markcite{cowie88}Cowie\ 1988). 

A star formation rate of 
$70\ {\rm h_{65}^{-2}\ M_\odot\ yr^{-1}\ Mpc^{-3}}$ for $q_0=0.5$
could produce a spheroid of mass 
$6\times 10^{10}\ {\rm h_{65}^{-2}\ M_\odot}$ in 0.8\ Gyr. 
Recent CO observations of two high redshift submm sources by
\markcite{frayer98}\markcite{frayer99}Frayer et al.\ (1998, 1999)
show that each source has enough molecular gas to form an $L^*$ galaxy.
Thus, with a duty cycle that self-consistently matches the galaxy
number density, the submm population could plausibly evolve into
the present-day spheroidal population.

In summary, we have reported new 850-$\mu$m source counts from wide-area SCUBA 
imaging data of the Hawaii Deep Survey fields SSA13, SSA17, and SSA22
and improved counts from deep SCUBA imaging data of SSA13 and the 
Lockman Hole. Our data rule out models that assume the submm flux is due 
only to dust in optical galaxies because such models vastly underpredict the 
observed counts. We find that our cumulative counts, $N(>S)$, above a
$3\sigma$ detection threshold of 2\ mJy follow a simple power-law behavior.
By introducing an additional parameter constrained by the EBL,
we extrapolate our cumulative counts below 2\ mJy and infer
that the bulk of the submm EBL resides in sources near 1\ mJy, in
agreement with a fluctuation analysis by 
\markcite{hughes98}Hughes et al.\ (1998) and lensed source
counts from \markcite{blain99}Blain et al.\ (1999).
The submm sources are plausible progenitors of the present-day 
spheroidal population.

\smallskip
\acknowledgments
We thank an anonymous referee for helpful suggestions, and we thank
Stephen Eales and Simon Lilly for communicating to us their
flux sensitivities and area coverages.
AJB acknowledges support from NASA through contract number P423274 from the
University of Arizona, under NASA grant NAG5-3042.

\end{document}